\documentclass[aps,prl,twocolumn,nofootinbib,showpacs]{revtex4}
\usepackage{graphicx}
\usepackage{amsfonts}
\usepackage{amssymb}
\usepackage{epsfig}
\usepackage{psfrag}
\usepackage{dsfont}
\usepackage{bm}
\newcommand{\beqn}{\begin{eqnarray}}
\newcommand{\eeqn}{\end{eqnarray}}
\newcommand{\be}{\begin{equation}}
\newcommand{\ee}{\end{equation}}

\def\beq{\begin{equation}}
\def\eeq{\end{equation}}
\def\beqn{\begin{eqnarray}}
\def\eeqn{\end{eqnarray}}

\def\s1{$s_{\alpha}$}
\def\s2{$s_{\gamma}$}
\def\s3{$s_{\delta}$}
\def\c1{$c_{\alpha}$}
\def\c2{$c_{\gamma}$}
\def\c3{$c_{\delta}$}
\def\s{Stueckelberg~}

\newcommand{\mathsym}[1]{{}}

\begin{document}
\begin{center}
\end{center}

\title{
Probing a Very Narrow  $Z'$ Boson with CDF and D0 Data}

\author{ Daniel Feldman,  Zuowei Liu and  Pran Nath}

\affiliation{Department of Physics, Northeastern University,
 Boston, MA 02115, USA \\\rm{(Received by \rm PRL 6 March 2006; Revised 24 May 2006; Published in \rm PRL 14 July 2006)}}

\begin{abstract}
The CDF and D0 data of nearly $475$ $ \text{pb}^{-1}$ in the
dilepton channel is used to probe a recent class of models,
Stueckelberg extensions of the Standard Model (StSM), which predict
a $Z'$ boson whose mass is of topological  origin  with a  very
narrow decay width. A Drell-Yan analysis for dilepton production via
this $Z'$  shows that  the current data put constraints on the
parameter space of the StSM. With a total integrated luminosity of
$8 $ $ \text{fb}^{-1}$, the very narrow  $Z'$ can be discovered up
to a mass of about 600 GeV. The StSM $Z'$ will be very distinct
since it can occur in the region where a  Randall-Sundrum graviton
is excluded.

\end{abstract}
\maketitle {\it  Introduction. $-$} In this Letter we investigate
the implications of the cumulative CDF  \cite{cdfdata} and D0
\cite{Abazov:2005pi}   data in the dilepton channel to
 probe the very narrow $Z'$ boson that arises in the $U(1)_X$ Stueckelberg extension of the Standard Model
 (StSM) \cite{Kors:2004dx}.
Thus  string models involving dimensional reduction and intersecting
D branes \cite{ghi} allow for the possibility of an abelian  gauge
boson gaining mass without the benefit of a Higgs phenomenon via the
Stueckelberg mechanism where the mass parameter is topological in
nature  \cite{ABL}. Indeed the Stueckelberg couplings have played an
important role in the D brane model building \cite{Ibanez:2001nd}.
The  topological mass scale can be obtained from dimensional
reduction and is typically the  size of the compactification scale
\cite{ghi}.  However, it could also be taken as an independent
parameter \cite{Ibanez:1998qp}. The model of Ref. \cite{Kors:2004dx}
involves a non-trivial mixing of the Stueckelberg and the Standard
Model (SM)  sectors via an additional term $
\mathcal{L}_{\text{St}}$ in the low energy effective Lagrangian so
that \be\mathcal{L}_{\text{St}}=-\frac{C_{\mu  \nu }C^{\mu\nu }
}{4}+g_XC_{\mu }\mathcal{J}_X^{\mu } -\frac{1}{2}\left(\partial
_{\mu }\sigma +M_1C_{\mu}+M_2B_{\mu }\right)^2\ee where $C_{\mu}$ is
the gauge  field for $U(1)_X$ and $\mathcal{J}_X^{\mu }$ gives
coupling to the hidden sector (HS) but has no coupling to the
visible sector (VS),
  $B_{\mu}$ is the gauge field associated with
$U(1)_Y$, $\sigma$ is the axion,  and $M_1$ and $M_2$ are  mass parameters  that appear in the
Stueckelberg extension.  After electroweak symmetry breaking  with a single Higgs doublet,
the gauge group  $SU(2)_L\times U(1)_Y\times U(1)_X$  breaks down to $U(1)_{em}$,  and the  neutral sector is
modified due to mixing with the Stueckelberg sector. The mass$^2$ matrix in the neutral sector
is a $3\times 3$ matrix and in the basis  $\left(C^{\mu }, B^{\mu },  A^{3\mu}\right)$ is given by
\be
M_{\text{St}}^2=
\left(
\begin{array}{ccc}
    M_1^2             & M_1M_2                                & 0 \\
    M_1M_2         & \frac{1}{4}v^2g_Y^2+M_2^2  & -\frac{1}{4}v^2g_2g_Y \\
      0                   & -\frac{1}{4}v^2g_2g_Y            & \frac{1}{4}v^2g_2^2
\end{array}
\right),
\ee
where $g_2(g_Y)$ are the gauge couplings in the $SU(2)_L(U(1)_Y)$  sectors, and $v = \langle H\rangle$ where $H$ is the
SM Higgs field.
${M}_{\text{St}}^2$ being  real and symmetric is diagonalized by an  orthonormal matrix ${O}$  so that
${O}^TM^2_{St} {O} = M^2_{\text{St-diag}}$ with the useful parameterization
\be {O}=\left(
\begin{array}{ccc}
 c_ {\psi }c_ {\phi} -s_ {\theta} s_ {\phi}s_ {\psi } & -s_ {\psi }c_ {\phi} -s_ {\theta} s_ {\phi} c_ {\psi } & -c_ {\theta} s_ {\phi}  \\
 c_ {\psi }s_{\phi} +s_ {\theta} c_ {\phi} s_ {\psi } & -s_ {\psi }s_ {\phi} +s_ {\theta} c_ {\phi} c_ {\psi } & c_ {\theta}c_ {\phi}  \\
 -c_ {\theta} s_ {\psi } & -c_ {\theta} c_ {\psi } & s_ {\theta}
\end{array}
\right).\ee
One then finds $t_{\phi}={M_2}/{M_1}$,   $t_
{\theta}={g_Y}c_{\phi}/g_2$ and $t_{\psi}= t_{\theta}t_{\phi} M_W^2
(c_{\theta}\left(M_{Z^{{\prime }}}^2-M_W^2\left(1+
t^2_{\theta}\right)\right))^{-1}$ where $s_{\theta}=\sin\theta,
c_{\theta}=\cos\theta, t_{\theta}= \tan\theta$, etc. Eq. (2)
contains one massless state, i.e.,  the photon, and two massive
states, i.e., the $Z$ and  $Z'$. The photon field here  is a linear
combination of $C^{\mu}, B^{\mu}, A^{3\mu}$ which distinguishes it
from other class of extensions [see, e.g.,
\cite{leike,Carena:2004xs,Cvetic:1995rj}], and in addition the model
contains a very narrow $Z'$ resonance. The effects of the
Stueckelberg  extension are contained in the parameters  $\epsilon
\equiv M_2/M_1$ and $M_1$. In the limit $\epsilon \to 0$ the
Stueckelberg sector decouples from the Standard Model.

{\it Electroweak constraints. $-$}
To determine  the  allowed corridors  in $\epsilon$ and $M_1$,
we follow a similar approach as  in the analysis of Refs. \cite{Nath:1999fs,marciano}   used in constraining  the size of extra dimensions.
  We begin by recalling that in the
 on-shell scheme the  $W$  boson mass including loop corrections is  given by \cite{Sirlin:1983ys}
  \be
 M_W^2=\frac{\pi \alpha}{ \sqrt 2 G_F \sin^2\theta_W (1-\Delta r)}  ,
 \label{wmass}
  \ee
 where the Fermi constant $G_F$ and the
fine structure constant $\alpha$  (at  $Q^2=0$)  are  known to a high degree  of accuracy.
The quantity $\Delta r$ is the radiative correction and is determined so that
 $\Delta r= 0.0363\pm 0.0019$ \cite{:2005em}, where the  uncertainty comes from error in the top mass and
 from  the error in $\alpha (M_Z^2)$.  Now since in the on-shell
 scheme $\sin^2\theta_W= (1-M_W^2/M_Z^2)$ one may use Eq. (\ref{wmass})  and the current experimental
 value of $M_W=80.425 \pm 0.034$ \cite{:2005em}  to make a prediction of
 $M_Z$.  Such a prediction within SM  is in excellent agreement with the
 current experimental value of $M_Z=91.1876\pm 0.0021$.
 Thus the above analysis requires that the effects of the Stueckelberg extension on the $Z$ mass must be
 such that they lie in the error corridor of the SM  prediction. We now calculate the
 error $\delta M_Z$   in the SM  prediction of $M_Z$ in order to limit $\epsilon$.  From Eq.  (\ref{wmass}) we find that
 $\delta \equiv\delta M_Z/M_Z|_{SM}$ is given by
  \be
\delta  = \sqrt{ \left(\frac{1-2\sin^2\theta_W}{\cos^3\theta_W} \frac{\delta M_W}{M_Z}\right )^2
 +\frac{\tan^4\theta_W (\delta \Delta r)^2}{ 4(1-\Delta r)^2 }  }.
 \label{delta1}
 \ee
 From Eq.  (2) the  Stueckelberg correction to the $Z$ mass in the
 region $M_1^2 \gg M_Z^2$ is given by
 $ |\Delta M_Z/M_Z|= \frac{1}{2} \sin^2\theta_W (1-{M_Z^2}/{M_1^2})^{-1} \epsilon^2$.
 Equating this  shift to the result of Eq. (5) one finds an upper bound on $\epsilon$
 \beqn
 |\epsilon| \lesssim0.061  \sqrt{ 1-(M_Z/M_1)^2} \label{epsilon} .
 \eeqn

 Next we  obtain in an  independent way the  constraint on $\epsilon$ by using  a fit to a standard set of electroweak
 parameters.    We follow closely the analysis of the LEP Working Group  \cite{:2005em} [see also Refs.   \cite{Baur:2001ze,Bardin:1999gt}],
 except that we will use the vector ($v_f$) and axial vector ($a_f$) couplings  for the fermions in the
  StSM.
Here, we exhibit as an example, the $Z$  couplings of the charged leptons in the StSM
\beqn
 v_{\ell }  (a_{\ell})   =\sqrt{\rho _{\ell }}(T_{3,\ell } \beta _L- Q_{\ell }(\beta_L\pm\beta _R)\kappa _{\ell } s^2_W),
 \eeqn
where $\beta_{L,R}$ are  as defined in Ref. \cite{Kors:2004dx}, and
 where $\rho_{\ell}$ and $\kappa_{\ell}$ (in general complex valued quantities) contain radiative corrections from propagator self-energies
and flavor specific vertex corrections and are as defined in Refs. \cite{Erler:2004nh,:2005em}.
The  SM limit corresponds to $\epsilon \to 0$, and  $\beta _{L,R}  \to1$.

\begin{table}[h]
\caption{\label{tab:table1} Results of the StSM fit to a standard
set of electroweak observables  at the $Z$ pole for $\epsilon$ in
the range  $(0.035-0.057)$ for $M_1=250$ GeV. The Pulls are
calculated as shifts from the SM fit via $\Delta {\rm Pull}=({\rm SM
- StSM})/\delta {\rm exp}$ and
  Pull(StSM)=Pull(SM)+ $\Delta$Pull.
 The  data in column 2 are taken from  Ref. \cite{pdg}.}
\begin{ruledtabular}
\begin{tabular}{cccccc}
Quantity                       & Value  (Experiment)                     &StSM        &$\Delta$Pull        \\
\hline
$\Gamma_Z$ [GeV]           & 2.4952  $\pm$ 0.0023      &(2.4948-2.4935)       &(0.4, 0.9)        \\
$\sigma_{had}$ [nb]        & 41.541  $\pm$ 0.037       &(41.478-41.481)       &(-0.1, -0.1)      \\
$R_e$                     & 20.804  $\pm$ 0.050       &(20.743-20.742)       &(-0.1, -0.2)        \\
$R_\mu$                    & 20.785  $\pm$ 0.033       &(20.744-20.743)       &(0.1, 0.2)        \\
$R_\tau$                   & 20.764  $\pm$ 0.045       &(20.791-20.790)       &(0.0, 0.1)        \\
$R_b$                      & 0.21643 $\pm$ 0.00072     &(0.21583-0.21583)     &(0.0, 0.0)        \\
$R_c$                      & 0.1686 $\pm$ 0.0047       &(0.1723-0.1723)       &(0.0, 0.0)        \\
$A^{(0,e)}_{FB}$           & 0.0145 $\pm$ 0.0025       &(0.0167-0.0174)       &(-0.2, -0.5)      \\
$A^{(0,\mu)}_{FB}$         & 0.0169  $\pm$ 0.0013      &(0.0167-0.0174)       &(-0.3, -0.9)      \\
$A^{(0,\tau)}_{FB}$        & 0.0188  $\pm$ 0.0017      &(0.0167-0.0174)       &(-0.3, -0.7)      \\
$A^{(0,b)}_{FB}$           & 0.0991  $\pm$ 0.0016      &(0.1046-0.1068)       &(-0.9, -2.2)     \\
$A^{(0,c)}_{FB}$           & 0.0708  $\pm$ 0.0035      &(0.0748-0.0764)       &(-0.3, -0.7)      \\
$A^{(0,s)}_{FB}$           & 0.098  $\pm$ 0.011      &(0.105-0.107)         &(-0.1, -0.3)      \\
$A_e$                      & 0.1515 $\pm$ 0.0019       &(0.1492-0.1524)       &(-1.0, -2.7)      \\
$A_\mu$                    & 0.142  $\pm$ 0.015        &(0.149-0.152)         &(-0.1, -0.3)      \\
$A_\tau$                   & 0.143  $\pm$ 0.004        &(0.149-0.152)         &(-0.5, -1.3)      \\
$A_b$                      & 0.923  $\pm$ 0.020        &(0.935-0.935)         &(0.0, 0.0)        \\
$A_c$                      & 0.671  $\pm$ 0.027        &(0.668-0.668)         &(0.0, 0.0)        \\
$A_s$                      & 0.895 $\pm$ 0.091         &(0.936-0.936)         &(0.0, 0.0)        \\

\end{tabular}
\end{ruledtabular}
 \end{table}

  Using the above modifications we have carried out a fit in the electroweak sector.
   Results of the  analysis are given in  Table 1 for  $M_1=250$ GeV and
  $\epsilon$ in the range (0.035-0.057) where the upper limit corresponds to Eq. (6) and the lower  limit yields $|\Delta{\rm Pull}| <1$.
  To indicate the quality of the fits we compute $\chi^2/$DOF= (20.1,16.2,18.4)/18 for $\epsilon =(0.057, 0.035,0.0)$ excluding $A_{FB}^{(0,b)}$ and
 $\chi^2/$DOF= (43.3,28.0,25.0)/19  including   $A_{FB}^{(0,b)}$ (where DOF represents degrees of freedom).
 We note that $\epsilon =0.035$ gives the same excellent fit  to the data as $\epsilon =0$ [SM\cite{:2005em}] case including or  excluding  $A_{FB}^{(0,b)}$.
For   $\epsilon =0.057$ the fit excluding  $A_{FB}^{(0,b)}$ is as
good as  for the SM case, but less so when one includes
$A_{FB}^{(0,b)}$. However, as is well  known  $A_{FB}^{(0,b)}$  is
also problematic in SM since it has  a large Pull.
 Thus Ref. [14] quotes the Pull for $A_{FB}^{(0,b)}$   in the range [-2.5,-2.8] and states  that the large shift could be due  to
 a fluctuation  in one or  more of the input measurements in their experimental fits. It is also stated  in Ref. \cite{Erler:2004nh}
 that at least some of the problem here may be experimental.  Thus it would appear that the determination of
  $A_{FB}^{(0,b)}$ is  on a somewhat less firm footing than  the other electroweak parameters.

The  Stueckelberg extension of the Standard Model is among a class of models where such an extension
can occur. Other examples are provided by the extension
$SU(2)_L\times U(1)_{R}\times U(1)_{B-L}\times U(1)_X$, or by the extension of the more popular
$SU(2)_L\times SU(2)_R\times U(1)_{B-L}$ Left-Right (LR) model  \cite{Mohapatra:1979ia} to give the gauge group
$SU(2)_L\times SU(2)_R\times U(1)_{B-L}\times U(1)_X$ (StLR).  Here the mixing matrix is still a consequence of
Eq. (1) except that $B_{\mu}$ now stands for the $U(1)_{B-L}$ gauge field.  The vector mass$^2$ matrix
in this case is $4\times 4$ involving the fields
$\left(
 C_{\mu },
 B_{\mu },
 A_{\mu  L}^3,
 A_{\mu  R}^3
\right)$.
The mass$^2$ matrix  leads to one massless state and  three  massive states  $Z, Z', Z''$.
It is easily checked that  the electro-magnetic  interaction  is given by
$\mathcal{L}_{\text{EM}}= e A_{\mu }^{\gamma} \left(\mathcal{J}_{B -L}^{\mu }+ \mathcal{J}_{2L}^{3 \mu }+ \mathcal{J}_{2R}^{3
\mu }\right)$
where
\be
\frac{1}{e^2} =\frac{1}{g^2} \left(1-\frac{M_2^2}{M_1^2}\right) + \frac{1}{g_Y^2} \left(1+\frac{M_2^2}{M_1^2}\right)
\ee
 and where $g_Y$  is related to $g=g_{2L} =g_{2R}$ and $g'$  by
$ {1}/{g_Y^2} ={1}/{g^2}+{1}/{g'^2}$. The above relations limit to
the standard LR relation as
 ${M_2}/{M_1}\to 0 $.
Quite  remarkably  the
$Z'$ couplings of
StLR  are very close to the $Z'$ couplings of StSM   and thus we will focus  the analysis on
StSM  and the results  for the StLR will be very similar.
\begin{figure}[t]
\vspace*{.2in}
\hspace*{-.2in}
\centering
\includegraphics[width=8cm,height=7cm]{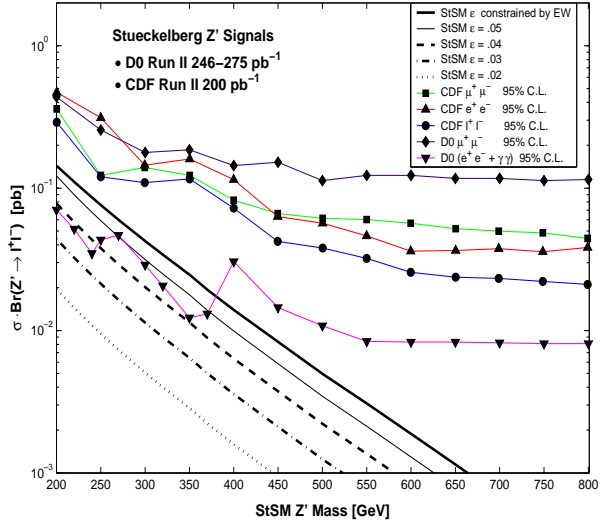}
\caption{  $Z'$ signal in StSM using the  CDF  \cite{cdfdata} and D0
\cite{Abazov:2005pi}
 data.
 The data puts a lower limit of about  250 GeV on
 $M_{Z'}$ for  $\epsilon \approx 0.035$ and 375 GeV for
 $\epsilon \approx 0.06$.}
\label{data}
\end{figure}

{\it Drell-Yan analysis of Stueckelberg $Z'.-$}
Next we discuss the production of the narrow $Z'$  by the Drell-Yan process at the Tevatron.
For the hadronic process $A +B \to  V +X$,  and the partonic subprocess $q  \bar{q}\to V \to l^+l^- $,
the dilepton production differential cross section to leading order (Born) is given by
\beqn
\frac{d \sigma _{AB}}{d M^2}=\frac{1}{s}  \sum _{q  }
\sigma^{q\bar{q}}\left(M^2\right)\mathcal{W}_{\left\{\text{AB}\left(q\bar{q}\right)\right\}}(\tau
),\hspace{.3cm}\tau  = \left.M^2\right/s
\nonumber\\
\mathcal{W}_{\left\{\text{AB}\left(q\bar{q}\right)\right\}}(\tau )=\int _0^1\int _0^1 d x d y \delta (\tau - x y)\mathcal{P}_{\left\{\text{AB}\left(q\bar{q}\right)\right\}} (x,y),
 \nonumber\\
\mathcal{P}_{\left\{\text{AB}\left(q\bar{q}\right)\right\}} (x,y)=f_{q,A}(x)f_{\bar{q},B}(y)+f_{\bar{q},A}(x)f_{q,B}(y).\nonumber
\eeqn
Here  $f_{q,A}$ and $f_{\bar q, A}$ are parton distribution functions (PDFs).
$\sigma^{q\bar q}$  is given in  \cite{Kors:2004dx}.
$\frac{d \sigma _{p\bar p}}{d M^2}$
 may be  calculated via a perturbative expansion in the strong coupling, $\alpha_s$, which is conventionally
absorbed into the Drell-Yan $K$ factor as discussed in detail in  Refs. \cite{Hamberg:1990np,Carena:2004xs,leike,Baur:2001ze}.


%
\begin{figure}[t]
\vspace*{.2in}
\hspace*{-.2in}
\centering
\includegraphics[width=8cm,height=7cm]{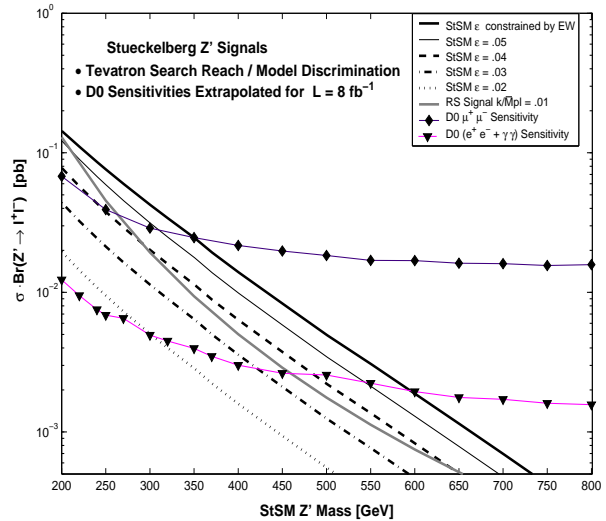}
\caption{ $Z'$ signal in StSM with 8 fb$^{-1}$ of data using an
extrapolation of the sensitivity of the  D0 \cite{Abazov:2005pi}
detector for the $\mu^+ \mu^-$ and $e^+ e^-+\gamma\gamma$ modes.
  The data will  put a lower limit of about 600 (300)  GeV  on  $M_{Z'}$ mass  for $\epsilon =0.06 (0.02)$.
  Also plotted for comparison is $\sigma\cdot Br(G\to l^+ l^-)$ for the RS case.}
\label{ig3}
\end{figure}

In Fig. 1 we give an analysis of the Drell-Yan  cross section for
the process $p\bar p\to Z' \to l^+ l^-$ as a function of  $M_{Z'}$.
The analysis is done at $\sqrt s = 1.96$ $\text{}\text{\rm{TeV}}$,
using the CTEQ5L \cite{Pumplin:2002vw}
 PDFs with a flat $K$ factor of $1.3$ for the appropriate comparisons with other models and with the CDF \cite{cdfdata} and D0 \cite{Abazov:2005pi} combined data in the dilepton channel.
 Remarkably one finds that the Stueckelberg $Z'$ for the case
$\epsilon \approx 0.06$  is eliminated up to about 375 GeV with the
 current data  (at 95\% C.L.). This lower limit decreases as $\epsilon$ decreases  but the current data  still constrain the model
 up to $\epsilon\approx 0.035$.  This result is in contrast to the LR,  $E_6$, and to the  little
 Higgs  models \cite{Han:2003wu}  where the   $Z'$ boson  has
already been eliminated up to  $(610 - 815)$  GeV with the CDF
\cite{cdfdata} and D0 \cite{Abazov:2005pi}  data. In Fig. 2 we give
the analysis of the discovery limit for the Stueckelberg $Z'$ with
an integrated  luminosity of $8$ $\text{fb}^{-1}$. Here we have
extrapolated the experimental sensitivity curves for the $\mu^+
\mu^-$ and for the more  sensitive $e^+ e^-+\gamma\gamma$ channel
downwards by a factor of $1/\sqrt N$ where $N$ is the ratio of the
expected integrated  luminosity to the current integrated
luminosity.  The analysis shows that a Stueckelberg $Z'$ can be
discovered up to a mass of about  600 GeV and if no effect is seen
one can put a lower limit  on the $Z'$  mass
 at about 600 GeV.  In Fig. 3 we give the exclusion plots in the $\epsilon -M_{Z'}$ plane using the current
 data and also using the total integrated luminosity of $8$ fb$^{-1}$ expected at the Tevatron.
 An analysis including hidden sector with $\Gamma_{\rm HS}=\Gamma_{\rm VS}$ is also exhibited.
  The exclusion plots show that even the hidden sector
 is beginning to be  constrained and  these constraints will  become even more severe with future data.
\begin{figure}[t]
\vspace*{.2in}
\hspace*{-.2in}
\centering
\includegraphics[width=8cm,height=8cm]{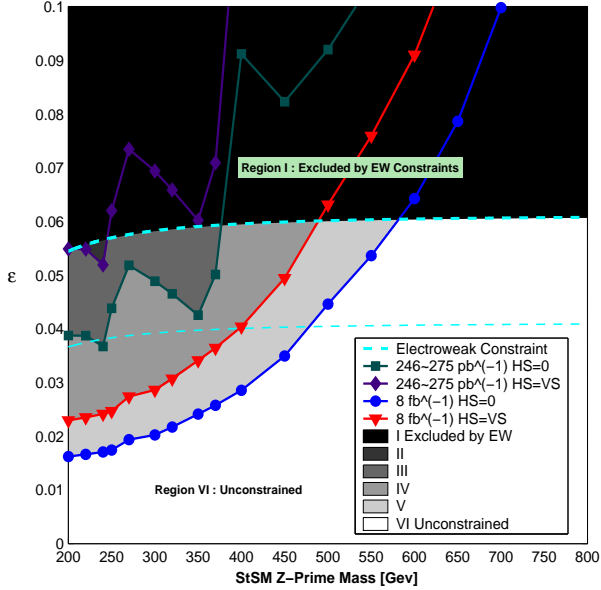}
\caption{ Exclusion plots in the $\epsilon -M_{Z'}$ plane  utilizing
the more sensitive D0  \cite{Abazov:2005pi}   $ e^+
e^-+\gamma\gamma$ mode with (a)   the  246-275 pb$^{-1}$ of data,
and (b)   8 fb$^{-1}$ of data  where an extrapolation of the
sensitivity curve is used.
  The upper  dashed curve
is the maximum value of $\epsilon$ allowed by Eq. (6) and the lower dashed curve corresponds to $|\Delta {\rm Pull}| <1$
(see the text for  the validity of imposing the lower constraint).
 Cases  with (without)
a hidden  sector are shown.  Regions  II, III, IV, and V are constrained by the conditions given at their
respective boundaries.}
\label{fig4}
\end{figure}

{\it Conclusion.$-$}
The type of $Z'$ boson that arises from the mixing of the Standard Model with the Stueckelberg
sector is very different from the $Z'$ bosons that normally arise in grand unified models \cite{leike} and in
 string models such as \cite{Cvetic:1995rj}, or in Kaluza-Klein excitations of the $Z$ in the compactifications of large
extra dimensions \cite{Antoniadis:1999bq}. The distinguishing feature is that
the decay width in the present case is exceptionally narrow with width $\leq 60$ MeV
for $M_{Z'}\leq 1$  TeV.
 It is interesting to note that there is a region of the parameter space where a Stueckelberg
$Z'$ boson may be mistaken for a narrow resonance of a
 Randall-Sundrum (RS)  \cite{Randall:1999ee}  warped geometry.
The RS warped geometry is a slice of anti$-–$de Sitter space
$(AdS_5)$ with the metric
$ds^2=exp(-2kr_c|\phi|)\eta_{\mu\nu}dx^{\mu}dx^{\nu} -r_c^2
d\phi^2$, $0\leq \phi\leq \pi$, where $r_c$  is the radius of the
extra dimension and  $k$  is the curvature of $AdS_5$. The overlap
of   $\sigma\cdot Br(Z'\to l^+l^-)$  and  $\sigma\cdot Br(G\to
l^+l^-)$ for the RS graviton is shown in Fig. 2 for the case $k/\bar
M_{\rm Pl}=0.01$ where $\bar M_{\rm Pl}= M_{\rm Pl}/\sqrt{8\pi}$ is
the reduced Planck mass. However, the constraints of the precision
electroweak data actually eliminate the RS graviton in this case
\cite{Davoudiasl:1999jd,Abazov:2005pi}. Thus if a resonance effect
is seen in the dilepton mass range of up to about 600 GeV in the CDF
and D0 data at the predicted level, the Stueckelberg $Z'$ would be \
a prime candidate since the RS graviton possibility is absent in
this case.

This research  was supported in part by NSF grant PHY-0546568. We
thank Darien Wood for many informative discussions related to
experiment and for a careful reading  of the
manuscript.\\\\\\\\\\\\\\\\\\


\begin{thebibliography}{999}


\bibitem{cdfdata}
 A.~Abulencia {\it et al.}  [CDF Collaboration],
  Phys.\ Rev.\ Lett.\  {\bf 95}, 252001 (2005)
  [arXiv:hep-ex/0507104].


\bibitem{Abazov:2005pi}
  V.~M.~Abazov {\it et al.}  [D0 Collaboration],
  Phys.\ Rev.\ Lett.\  {\bf 95}, 091801 (2005)
  [arXiv:hep-ex/0505018].

\bibitem{Kors:2004dx}
  B.~Kors and P.~Nath,
  Phys.\ Lett.\ B {\bf 586}, 366 (2004); 
  JHEP {\bf 0412}, 005 (2004); 
  JHEP {\bf 0507}, 069 (2005).

\bibitem{ghi}
D.~M.~Ghilencea, L.~E.~Ibanez, N.~Irges and F.~Quevedo,
JHEP {\bf 0208} (2002) 016; 
D.~M.~Ghilencea,
Nucl.\ Phys.\ B {\bf 648} (2003) 215.  

\bibitem{ABL}
T.~J.~Allen, M.~J.~Bowick and A.~Lahiri,
 Mod.\ Phys.\ Lett.\ A {\bf 6} (1991) 559.

\bibitem{Ibanez:2001nd}
  I.~Antoniadis, E.~Kiritsis and T.~N.~Tomaras,
  Phys.\ Lett.\ B {\bf 486}, 186 (2000)
  [arXiv:hep-ph/0004214];
  L.~E.~Ibanez, F.~Marchesano and R.~Rabadan,
JHEP {\bf 0111} (2001) 002; 
%
R.~Blumenhagen, V.~Braun, B.~K\"ors and D.~L\"ust,
hep-th/0210083;
  I.~Antoniadis, E.~Kiritsis, J.~Rizos and T.~N.~Tomaras,
  Nucl.\ Phys.\ B {\bf 660}, 81 (2003);
 C.~Coriano', N.~Irges and E.~Kiritsis,
  arXiv:hep-ph/0510332.

\bibitem{Ibanez:1998qp}
L.~E.~Ibanez, R.~Rabadan and A.~M.~Uranga,
Nucl.\ Phys.\ B {\bf 542}. 112  (1999).

\bibitem{leike}
  A.~Leike,
  Phys.\ Rept.\  {\bf 317}, 143 (1999).

\bibitem{Carena:2004xs}
  M.~Carena, A.~Daleo, B.~A.~Dobrescu and T.~M.~P.~Tait,
  Phys.\ Rev.\ D {\bf 70}, 093009 (2004).

\bibitem{Cvetic:1995rj}
  M.~Cvetic and P.~Langacker,
  Phys.\ Rev.\ D {\bf 54}, 3570 (1996).


\bibitem{Nath:1999fs}
P.~Nath and M.~Yamaguchi,
Phys.\ Rev.\ D {\bf 60}, 116004 (1999); 
M.~Masip and A.~Pomarol,
Phys.\ Rev.\ D {\bf 60}, 096005 (1999);
  R.~Casalbuoni, S.~De Curtis, D.~Dominici and R.~Gatto,
  Phys.\ Lett.\ B {\bf 462}, 48 (1999);
T.~G.~Rizzo and J.~D.~Wells,
Phys.\ Rev.\ D {\bf 61}, 016007 (2000); 
C.~D.~Carone,
Phys.\ Rev.\ D {\bf 61}, 015008 (2000).

\bibitem{marciano}
W.~J.~Marciano,
Phys.\ Rev.\ D {\bf 60}, 093006 (1999).

\bibitem{Sirlin:1983ys}
  A.~Sirlin,
  Phys.\ Rev.\ D {\bf 29}, 89 (1984);
  W.~J.~Marciano and A.~Sirlin,
  Phys.\ Rev.\ D {\bf 29}, 945 (1984).

\bibitem{:2005em}
    [ALEPH Collaboration],
  arXiv:hep-ex/0509008.

\bibitem{Baur:2001ze}
  U.~Baur, O.~Brein, W.~Hollik, C.~Schappacher and D.~Wackeroth,
  Phys.\ Rev.\ D {\bf 65}, 033007 (2002).

\bibitem{Bardin:1999gt}
  D.~Y.~Bardin et.al.,
  arXiv:hep-ph/9902452.

\bibitem{Erler:2004nh}
  J.~Erler and P.~Langacker,
  arXiv:hep-ph/0407097.


\bibitem{pdg}
  S.~Eidelman {\it et al.}  [Particle Data Group],
  Phys.\ Lett.\ B {\bf 592}, 1 (2004).

\bibitem{Mohapatra:1979ia}
  R.~N.~Mohapatra and G.~Senjanovic,
  Phys.\ Rev.\ Lett.\  {\bf 44}, 912 (1980).

\bibitem{Hamberg:1990np}
  R.~Hamberg, W.~L.~van Neerven and T.~Matsuura,
  Nucl.\ Phys.\ B {\bf 359}, 343 (1991).


\bibitem{Pumplin:2002vw}
  J.~Pumplin, D.~R.~Stump, J.~Huston, H.~L.~Lai, P.~Nadolsky and W.~K.~Tung,
  JHEP {\bf 0207}, 012 (2002).



\bibitem{Han:2003wu}
  T.~Han, H.~E.~Logan, B.~McElrath and L.~T.~Wang,
  Phys.\ Rev.\ D {\bf 67}, 095004 (2003)
  [arXiv:hep-ph/0301040].



\bibitem{Antoniadis:1999bq}
  I.~Antoniadis, K.~Benakli and M.~Quiros,
  Phys.\ Lett.\ B {\bf 460}, 176 (1999);
 P.~Nath, Y.~Yamada and M.~Yamaguchi,
  Phys.\ Lett.\ B {\bf 466}, 100 (1999).


\bibitem{Randall:1999ee}
  L.~Randall and R.~Sundrum,
  Phys.\ Rev.\ Lett.\  {\bf 83}, 3370 (1999).

\bibitem{Davoudiasl:1999jd}
  H.~Davoudiasl, J.~L.~Hewett and T.~G.~Rizzo,
  Phys.\ Rev.\ Lett.\  {\bf 84}, 2080 (2000).

\end{thebibliography}
\end{document}